%% file: JSAC_Zuleitav2.tex
\documentclass[10pt, A4paper, doublecolumn]{IEEEtran}
\usepackage{epsfig,graphics,psfrag,amsmath,float, subfig}
\usepackage{latexsym,amssymb,cite, algorithmic, algorithm}

\newtheorem{Thm}{\textbf{Theorem}}
\newtheorem{Lem}{\textbf{Lemma}}
\newtheorem{Def}{\textbf{Definition}}

\begin{document}
\title{Balancing Egoism and Altruism on the single beam MIMO Interference Channel}
\author{Zuleita~K.~M.~Ho, ~\IEEEmembership{Member,~IEEE,}
David~Gesbert \IEEEmembership{Member,~IEEE,}\thanks{This work has been performed in the framework of the European research
project SAPHYRE, which is partly funded by the European Union under 
its FP7 ICT Objective 1.1 - The Network of the Future.}}\maketitle
\markboth{\MakeLowercase{\textit{submitted to}} IEEE Transactions on Signal Processing}{}
\begin{abstract}
This paper considers the so-called multiple-input-multiple-output interference channel (MIMO-IC) which has relevance in applications such as multi-cell coordination in
cellular networks as well as spectrum sharing in cognitive radio networks among
others. We consider a beamforming design framework based on striking a compromise
between beamforming gain at the intended receiver (Egoism) and the mitigation of
interference created towards other receivers (Altruism). Combining egoistic and altruistic beamforming has been shown previously in several papers to be instrumental to optimizing the rates in a  multiple-input-single-output interference channel MISO-IC (i.e. where receivers have no interference canceling capability). Here, by using the framework of Bayesian games, we shed more light on these game-theoretic concepts in the more general context of MIMO channels and more particularly when coordinating parties only have channel state information (CSI) of channels that they can measure directly. This allows us to derive distributed beamforming techniques. We draw parallels with existing work on the MIMO-IC, including rate-optimizing and interference-alignment precoding techniques, showing how such techniques may be improved or re-interpreted through a common prism based on balancing egoistic and altruistic beamforming. Our analysis and simulations currently limited to single stream transmission per user attest the improvements over known interference alignment based methods in terms of sum rate performance in the case of so-called asymmetric networks.  

\end{abstract}
\section{Introduction}
The mitigation of interference in multi-point to multi-point radio systems is of utmost
importance and has relevance in several practical contexts. Among the more
popular cases, we may cite the optimization of multi cell multiple-input-multiple-output (MIMO) systems with full
frequency reuse and cognitive radio scenarios featuring two or more service
providers sharing an identical spectrum license on  overlapping coverage
areas. In all these cases, the system may be modeled as a network of $N_c$
interfering radio links where each link consists of a sender trying to
communicate messages to a unique receiver in spite of the interference arising
from or created towards other links. 

% Recently, the attention of the research
% community was drawn to the so-called {\em coordinated} transmission methods where
% interference effects are mitigated or even exploited in exchange
% for an additional overhead in exchanging data symbols and channel state 
% information (CSI) between the transmitters. Bringing coordination to its fullest
% (i.e. assuming a complete sharing of data and CSI), multiple transmit antennas
% can be exploited as a virtual MIMO array and optimal forms of precoding allow the
% system designer to effectively exploit
% interference \cite{Shamai2001,Jafar2002,Weingarten2004,Kaviani2008,Shim2007}. In contrast, in a scenario where
% the backhaul network cannot support a complete sharing of data symbols across all transmitters,  the channel then remains a so-called {\em
% interference-channel} whereby the senders can resort to a milder form of
% coordination that does not require joint encoding of data packets. 
% Coordination over the interference channel may take place over one or several domains characterizing the
% transmission parameters of each sender such as the choice of power levels \cite{Gesbert2008},
% beamforming vectors \cite{Kaviani2008,Shim2007,Ye2003,Ku2008,Tenenbaum2008},
% assigned subcarriers in OFDMA \cite{Prasad2009}, scheduling \cite{Kiani2008, Choi2006} etc to cite a few.

For system limitation or privacy reasons, when the backhaul network cannot support a complete sharing of data symbols across all transmitters (Txs), the channel remains an interference channel (IC). Coordination in terms of beamforming is required to be decentralized in the sense that global channel state information at transmitters (CSIT) may not be available everywhere. In the context of distributed beamforming, game theory appears as a sensible approach as a basis for algorithm design. Recently an interesting game theory framework for beamforming-based coordination was proposed for the multiple-input-single-output (MISO) case by which the transmitters (e.g. the base stations)
seek to strike a compromise between selfishly serving their users while ignoring the interference effects on the one
hand,  and altruistically minimizing the harm they
cause to other non-intended receivers on the other hand.  An important result in this area was the characterization of all so-called Pareto rate optimal beamforming solutions for the two-cell case in  the form of positive linear  combinations of the  purely selfish and purely
altruistic beamforming solutions \cite{Jorswieck2007,Jorswieck2008,Jorswieck2008a} and \cite{Lindblom2009,Lindblom2010,Lindblom2010a} in the case of partial CSI. Unfortunately, how or whether at all this analysis can be extended to the context of MIMO interference channels (i.e. where receivers have themselves multiple antennas and interference cancelling capability) remains an open question. 

In parallel, coordination on the MIMO interference channel has emerged as a very popular topic in its own right, with several important non-game related contributions shedding light on rate-scaling optimal precoding strategies based on so-called interference alignment, subspace optimization, alternated maximum signal to interference and noise ratio (SINR) optimization, \cite{Khandani2008, Jafar2008,Heath2009} and rate-maximizing precoding strategies \cite{Ye2003,Wolfgang2009}, to cite just a few examples.

Interference alignment based strategies exhibit the designed feature of rendering interference cancellable (when feasible, according to the available degrees of freedom) at both the transmitter and receiver side. Such a behaviour is optimal in the large signal to noise ratio (SNR) region when Rxs have single user decoder and interference is the key bottleneck. At finite SNR, various strategies exist which aim at maximizing a link quality metric individually over each link, while taking interference into account. This often takes the form of maximizing the link's SINR or minimizing minimum-mean-square-error (MMSE). This approach provides good rates in symmetric networks where all links are subject to impairments (noise, average interference) of similar level. In more general and practical situations however, we argue that a better sum rate may be obtained from a proper and different weighting of the egoistic and altruistic objective at each individual link.  This situation is particularly important when more links are subject to statistically stronger interference than others, a case which has so far received little attention and which we shall refer here as asymmetric networks.   For this purpose, we suggest to re-visit the problem of coordinated beamforming design by directly building on the game theoretic concept of egoistic and altruistic game equalibria. Because our focus is on scenarios where CSI is not fully available, we consider a class of games suitable to the case of partial information-based decision making, called Bayesian games. Note that this is different from the limited CSI feedback scenario studied by previous authors \cite{boelcskei2009} who consider channel quantization requirement as function of SNR. Our approach is two fold, first derive analytically the  game equalibria. Second, exploit the obtained equilibria solution into heuristic design of a practical beamforming teachnique. The behaviour of our solution is then studied both theoretically (large SNR regime) and tested by simulations.

More specifically in this paper, our contributions are as follows: 
\begin{itemize}

\item  We define the egoistic and altruistic objective functions and derive analytically the equilibria of so-called egoistic and altruistic Bayesian games \cite{Harsanyi}.
\item Based on the equilibria, we propose a practical distributed beamforming scheme which provides a game-theoretic interpretation of the distributed sum rate maximization problem the MIMO-IC, such as \cite{Wolfgang2009}. 
\item The proposed techniques allows a tradeoff between the reduced complexity/feedback and the rate maximization offered by \cite{Wolfgang2009}.
\item We show that our algorithm exhibits the same rate scaling (when SNR grows) as shown by recent interesting interference alignment based methods \cite{Khandani2008,Jafar2008,Heath2009} which operates on the same feedback assumption as the proposed beamforming scheme. At finite SNR, we show improvements in terms of sum rate, especially in the case of asymmetric networks where interference-alignment methods are unable to properly weigh the contributions on the different interfering links to maximize the sum rate.  This situation is particularly relevant. In practical contexts where for complexity limitation reasons only a subset of cells (links) is coordinated across, while other uncoordinated links contribute to additional unequal amounts of unstructural interference.
 
\end{itemize}

%%%%%%%%%%

\subsection{Notations}
The lower case bold face letter represents a vector whereas the upper case bold face letter represents a matrix. $(.)^H$ represents the complex conjugate transpose. $\mathbf{I}$ is the identity matrix. $V^{(max)}(\mathbf{A})$ (resp. $V^{(min)}(\mathbf{\mathbf{A}})$) is the eigenvector corresponding to the largest (resp. smallest) eigenvalue of $\mathbf{\mathbf{A}}$. $\mathcal{E}_{B}$ is the expectation operator over the statistics of the random variable $B$. $\mathbb{S} \setminus \mathbb{B}$ define a set of elements in $\mathbb{S}$ excluding the elements in $\mathbb{B}$. $Tr(\mathbf{A})$ denotes the trace of matrix $\mathbf{A}$.

 \input{chmodel}

\input{bayesian_games}

 \input{egosol_fb}

 \input{altsol_fb}

 \input{dbs_fb}

\section{Simulation Results}\label{section:result}
In this section, we investigate the sum rate performances of \emph{DBA} in comparison with several related methods, namely the \emph{Max-SINR} method \cite{Jafar2008}, the alternated-minimization  (\emph{Alt-Min}) method for interference alignment \cite{Heath2009} and the sum rate optimization method (\emph{SR-Max}) \cite{Wolfgang2009}.  The \emph{SR-Max} method is by construction optimal but is more complex and requires extra sharing or feedback of pricing information among the transmitters.  To ensure a fair comparison, all the algorithms in comparisons are initialized to the same  solution and have the same stopping condition. The algorithms are considered to reach convergence if the sum rates achieved between successive iterations have difference less than 0.001. We perform sum rate comparisons in both symmetric channels and asymmetric channels where links undergo different levels of out-of-cluster noise.  Define the Signal to Interference ratio of link $i$ to be $SIR_i= \frac{\alpha_{ii}}{ \sum_{j \neq i}^{N_c} \alpha_{ij}}$. The $SIR$ is assumed to be 1 for all links, unless otherwise stated. Denote the difference in SNR between two links in asymmetric channels by $\Delta SNR$. 
Note that the proposed algorithm is not limited to the following settings, but can be applied to network with arbitrary players and number of antennas.

\subsection{Symmetric Channels}

% First we consider the MISO case and highlight the Pareto optimality of the iteration in \eqref{eqt:update}. In Fig. \ref{fig:sumrate_sym_miso_15db}, the pareto boundary of a 2-link symmetric channel is illustrated. The trajectory of \emph{DBA-BF} reaching the Pareto Boundary is plotted and the convergence point assuming the fairness-based stop condition described in \eqref{eqt:stop_cond} is shown.  
% \begin{figure}
%   \begin{center}
%    \includegraphics[height=7cm, width=9cm]{sumrate_sym_miso_15db.eps}
%  \caption{\emph{DBA-BF} reaches the Pareto Boundary in a 2-link symmetric channel realization.}\label{fig:sumrate_sym_miso_15db}
%   \end{center}
%  \end{figure}

\begin{figure}
  \begin{center}
   \includegraphics[height=7cm, width=9cm]{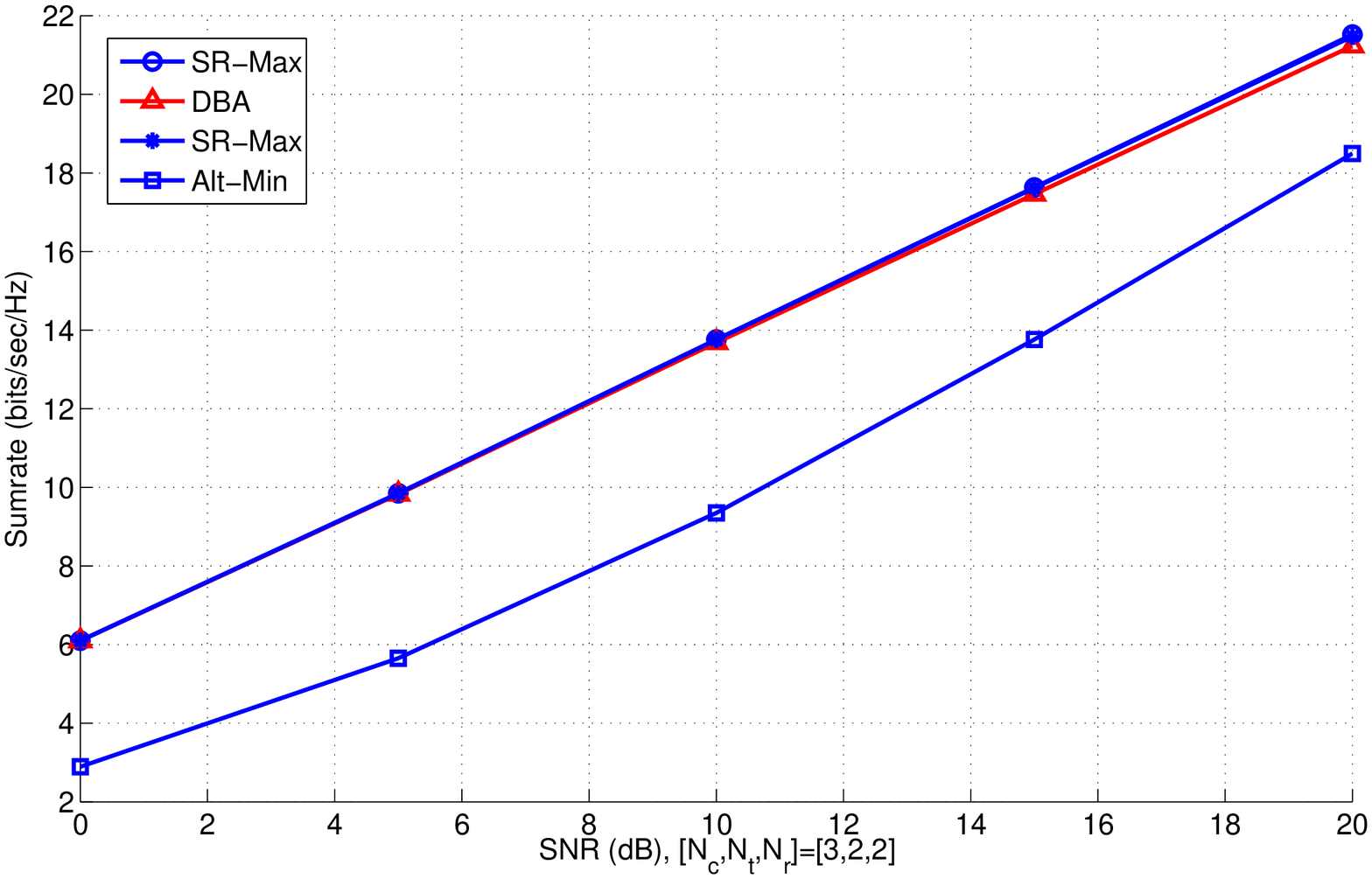}
 \caption{Sum rate comparison in multi links systems is illustrated with $[N_c,N_t,N_r]=[3,2,2]$ with increasing SNR. \emph{DBA}, \emph{SR-Max} and \emph{Max-SINR} achieve very close performance in symmetric networks.}\label{fig:sumrate_lam_nc3nt2nr2}
  \end{center}
 \end{figure}

Fig. \ref{fig:sumrate_lam_nc3nt2nr2} illustrates the sum rate comparison of \emph{DBA} with \emph{Max-SINR}, \emph{Alt-Min} and \emph{SR-Max} in a system of 3 links and each Tx and Rx have 2 antennas. Since interference alignment is feasible in this case, the sum rate performance of \emph{SR-Max} and \emph{Max-SINR} increases linearly with SNR. \emph{DBA} achieves sum rate performance with the same scaling as \emph{Max-SINR} and \emph{SR-Max} (i.e. multiplexing gain of 3).  Therefore these methods seem to perform similarly in symmetric channels. 

%  In Fig. \ref{fig:sumrate_nc5nt2nr2}, we show the sum rate in a system of 5 links where each Tx and Rx are equipped with 2 antennas. Note that in this case interference alignment is \emph{infeasible}. The sum rate performance saturates for all algorithms at high SNR regime. \emph{DBA-RF}, \emph{SR-Max} and \emph{Max-SINR} achieve very close performance to each other, but with less message exchange that in the \emph{SR-Max} technique.
%  
%  \begin{figure}
%   \begin{center}
%    \includegraphics[height=7cm, width=9cm]{sumrate_sym_nc5nt2nr2.eps}
%  \caption{Sum rate comparison in multi links systems with $[N_c,N_t,N_r]=[5,2,2]$ with increasing SNR. \emph{DBA-RF}, \emph{SR-Max} and \emph{Max-SINR} achieve very close performance.} \label{fig:sumrate_nc5nt2nr2}
%   \end{center}
%  \end{figure}
%  
 \subsection{Asymmetric Channels}
\begin{figure}[ht]
  \begin{center}
  \subfloat[Asymmetric uncontrolled interference]{\label{fig:asym_ch_model1}\includegraphics[width=9cm, height=6cm]{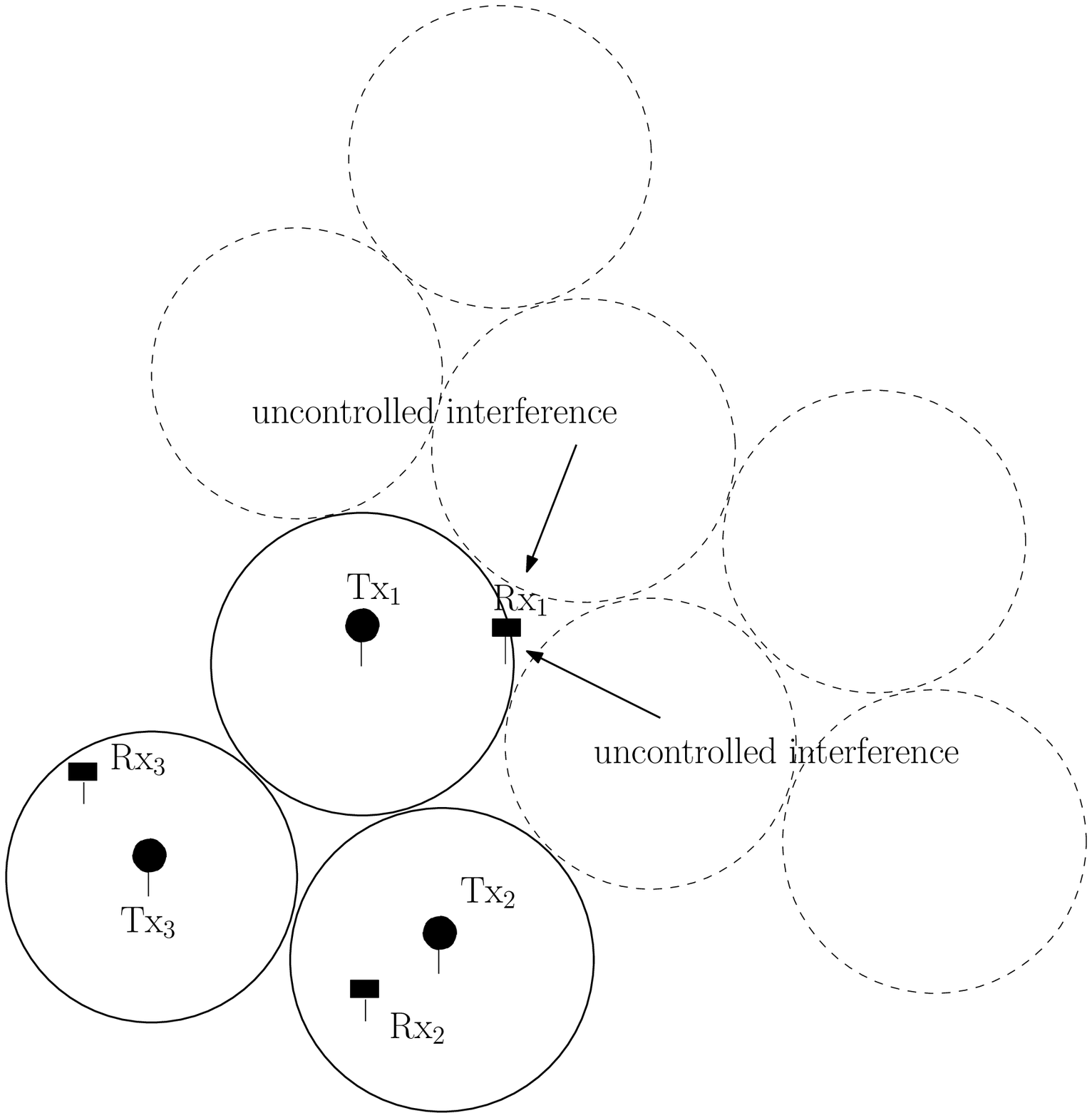}}  \\              
  \subfloat[Asymmetric uncontrolled interference and interference within cluster]{\label{fig:asym_ch_model2}\includegraphics[width=9cm, height=6cm]{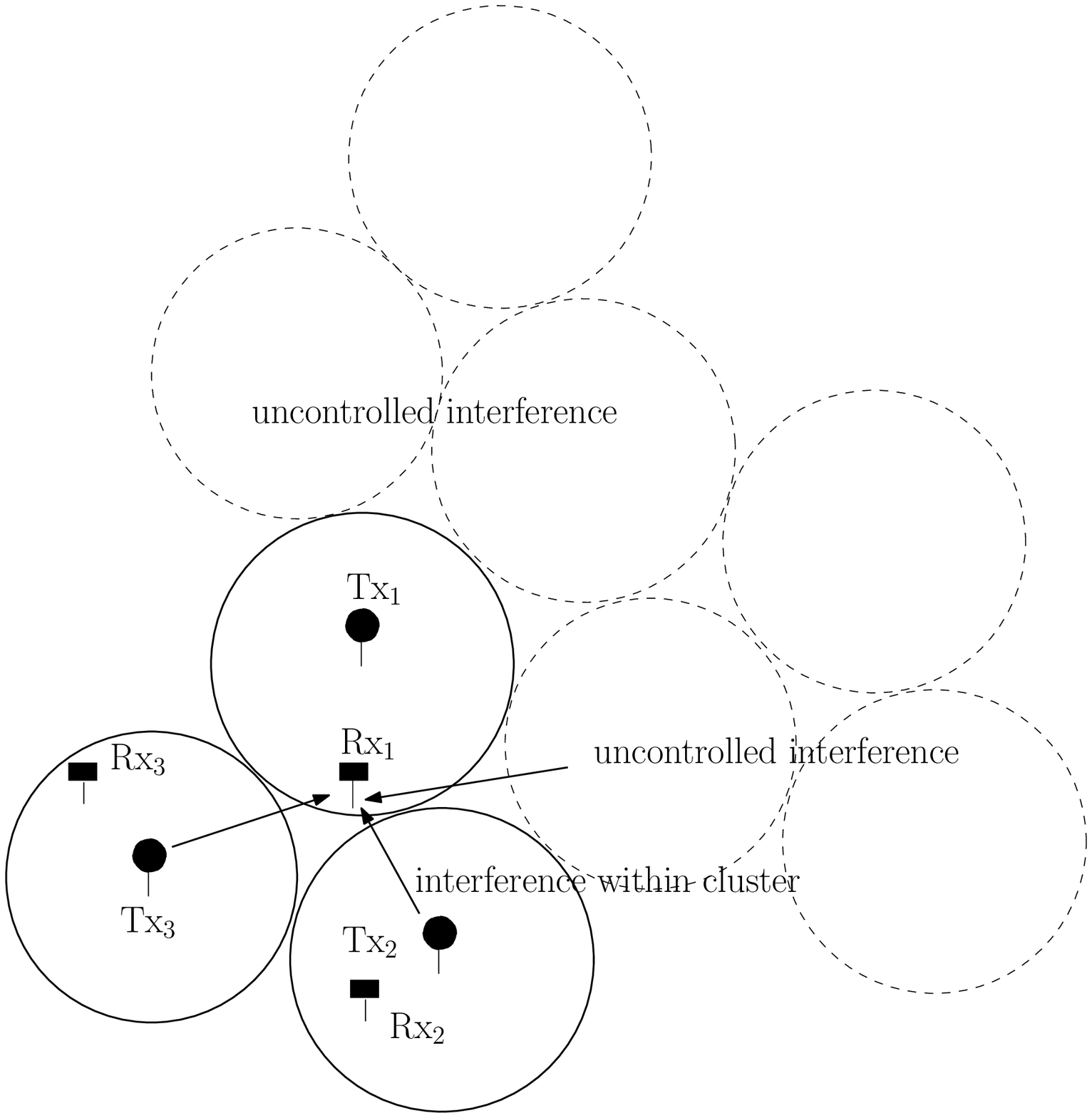}}\\
  \subfloat[Asymmetric desired channel power]{\label{fig:asym_ch_model3}\includegraphics[width=9cm, height=6cm]{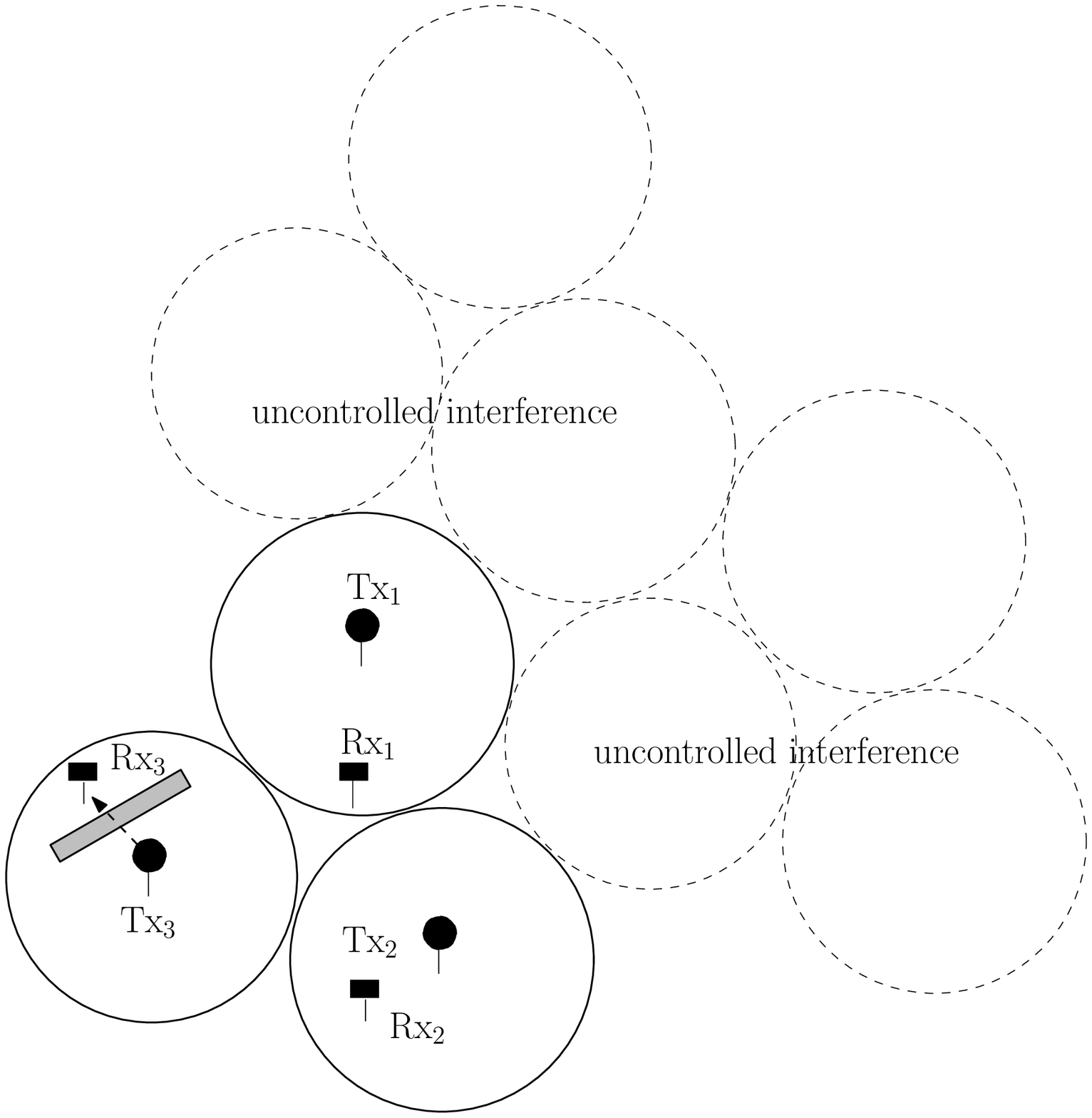}}
  \caption{Scenarios where asymmetry of channels are illustrated. Fig. \ref{fig:asym_ch_model1} illustrates asymmetry of uncontrolled interference. Fig. \ref{fig:asym_ch_model2} illustrates asymmetry of both uncontrolled interference and interference within cluster. Fig \ref{fig:asym_ch_model3} illustrates the asymmetric strength of desired channel power.}
  \label{fig:asym_ch_model}
  \end{center}
\end{figure}

In the asymmetric system, some links undergo uneven levels of noise and uncontrolled interference. Another aspect is that more links can experience greater path loss or shadowing than others. Here we consider a few typical scenarios for which could constitute asymmetric networks, as shown in Fig. \ref{fig:asym_ch_model}.
 
\subsubsection{Asymmetric uncontrolled interference power, illustrated in Fig. \ref{fig:asym_ch_model1}} 
\begin{figure}
  \begin{center}
   \includegraphics[height=7cm, width=9cm]{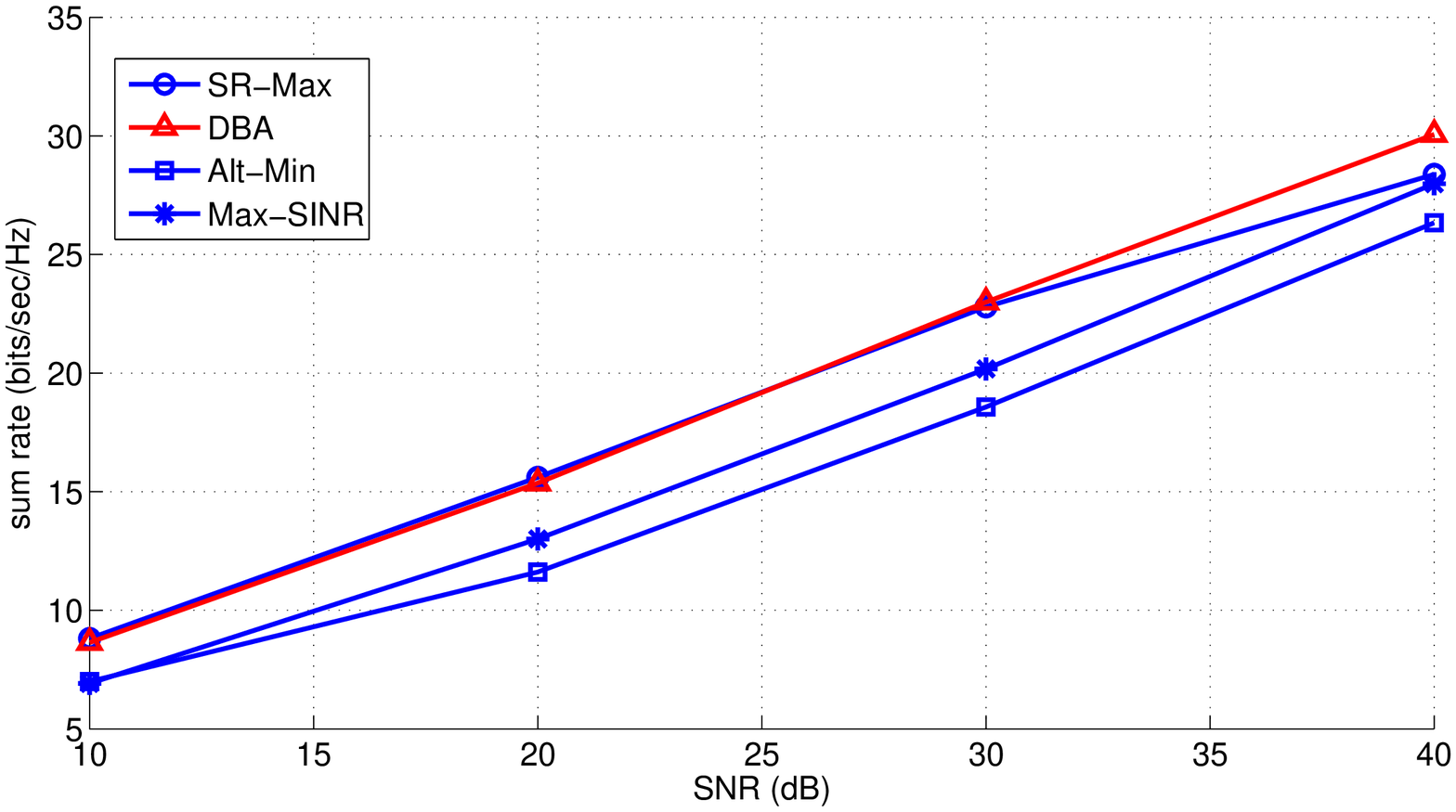}
 \caption{Sum rate performance for asymmetric channel, with one link under strong noise, is illustrated. The strong noise, from out of cluster interference, is 20dB stronger than other links. \emph{DBA} outperforms standard \emph{IA} methods thanks to a proper balance between egoistic and altruistic beamforming algorithm.}\label{fig:asym_noise}
  \end{center}
 \end{figure}

In Fig. \ref{fig:asym_noise}, there are 3 links in the system in which the noise and unstructural interference in one of the links are 20dB stronger than the other two links. This set up captures the scenario that one link is at the boundary of the coordination cluster and suffer from strong out-of-cluster noise. The SIR of every link is assumed to be 10 dB. in this scenario, \emph{DBA} outperforms interference alignment based methods because they are unable to properly weigh the importance of each link in the overall sum rate. \emph{SR-Max} is by construction sum rate optimal. However, in the asymmetric network, we observe by simulation that the convergence may require more iterations than other algorithms and the increment in sum rate per iteration can be small in some channel realizations. 

\subsubsection{Asymmetric uncontrolled interference power and interference within cluster, illustrated in Fig. \ref{fig:asym_ch_model2}} 
\begin{figure}
  \begin{center}
   \includegraphics[height=7cm, width=9cm]{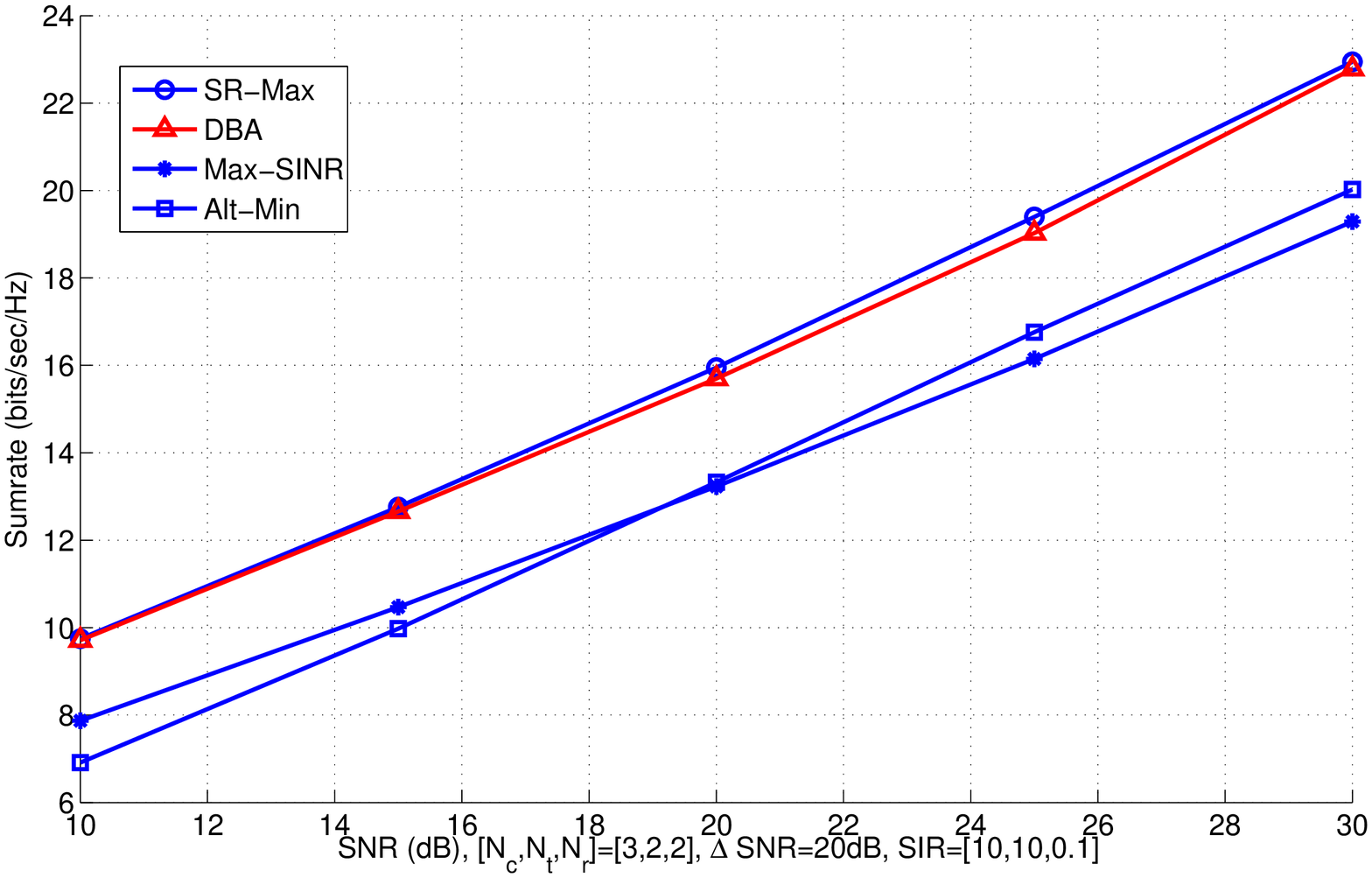}
 \caption{Sum rate performance for asymmetric channel, with one link under strong interference within the cooperating cluster, is illustrated.}\label{fig:asymmetric}
  \end{center}
 \end{figure}

In Fig. \ref{fig:asymmetric}, we compare the sum rate performance in the same set up as in Fig. \ref{fig:asym_noise}, except that the SIR's of the links are $[10,10,0.1]$ respectively. Thus,  link 3  not only suffers from strong out of cluster noise, but also suffers from strong interference within the cluster. The asymmetry penalizes the \emph{Max-SINR} and interference alignment methods because they are unable to properly weigh the contributions of the weaker link in the sum rate. The \emph{Max-SINR} strategy turns out to make link 3 very egoistic in this example, while its proper behavior should be altruistic.
In contrast, \emph{DBA} exploits useful statistical information, allowing weaker link to allocate their spatial degrees of freedom wisely towards helping stronger links  and vice versa, yielding a better sum rate for the same feedback budget. The performance is very close to \emph{SR-Max}, with less information exchange.

\subsubsection{Asymmetric desired channel power, illustrated in Fig. \ref{fig:asym_ch_model3}}
\begin{figure}
  \begin{center}
   \includegraphics[height=7cm, width=9cm]{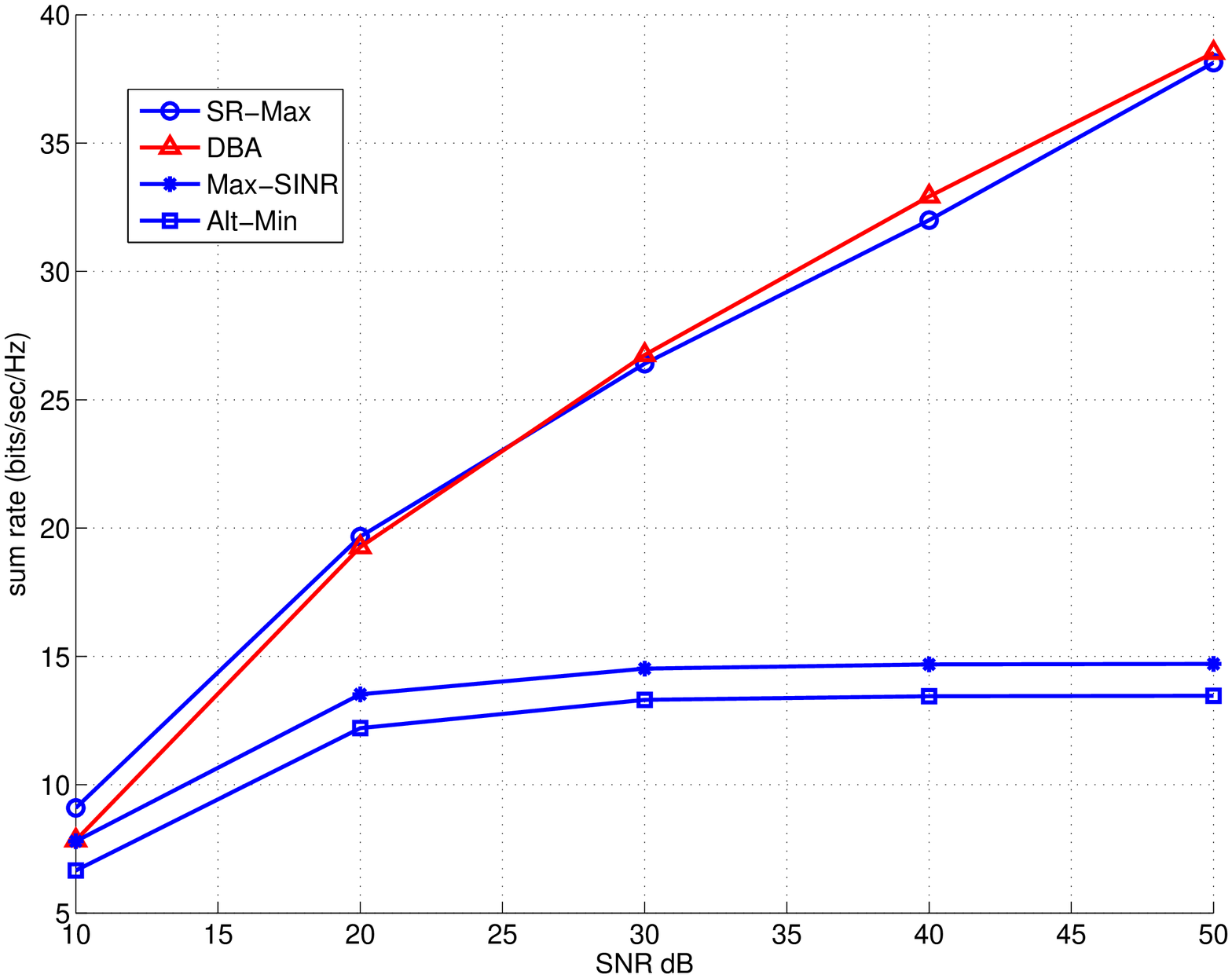}
 \caption{Sum rate performance for asymmetric channel  is illustrated. The direct channel gain of link 1 is 30dB weaker than other links. }\label{fig:weaklink}
  \end{center}
 \end{figure}

In Fig. \ref{fig:weaklink}, there are 3 links cooperating in the system. Each Tx and Rx has 2 antennas and has 1 stream transmission. The noise at each Rx is the same. The system is asymmetric in a sense that the direct channel gain $H_{11}$ of link 1 is 30dB weaker than other links in the network. This set up models a realistic environment where the user suffers strong shadowing. \emph{DBA} achieves sum rate closed to \emph{SR-Max} and much better than other interference alignment based schemes \emph{Max-SINR} and \emph{Alt-Min}.

 \section{Conclusion}
We model the distributed beamforming optimization problem on MIMO interference channel using the framework of Bayesian Games which allow players to have imcomplete information of the game, in this case the channel state information. Based on the incentives of the players, we proposed two games: the Egoistic Bayesian Game (players selfishly maximize its rate) and the Altruistic Bayesian Game (players altruisticly minimize interference generated towards other players). We proved the existence of equilibria of such games and the best response strategy of players are computed. Inspired from the equilibria, a beamforming technique based on balancing the egoistic and the altruistic behavior with the aim of maximizing the sum rate is proposed. Such beamforming algorithm exhibits the same optimal rate scaling (when SNR grows) shown by recent iterative interference-alignment based methods. The proposed beamforming algorithm acheives close to optimal sum rate maximization method \cite{Wolfgang2009} without additional pricing feedbacks from users and outperform interference alignment based methods in terms of sum rate in asymmetric networks.

\input{appendix}

\bibliographystyle{IEEEbib}
\bibliography{zuleitabib}

\end{document}

%% file: chmodel.tex
\section{Bayesian Games Definition on interference channel}\label{section:bayesian}
Let $\mathcal{N}= \left\{ 1 , \ldots, N \right\}$ be a set containing  a finite set $\mathcal{N}_c$, with cardinality $N_c \leq N $, of cooperating transmitters (Txs), also termed as players. From now on, we use players and Txs interchangably. We call the set $\mathcal{N}_c$ a coordination cluster and Txs outside the cluster will contribute to uncontrolled interference. The provided model has general applications in which the Txs can be base stations in cellular downlink where typically coordination is restricted to a subset of neighbouring cell sites while more distant sites cannot be coordinated over \cite{Gesbert2010} ; nodes in ad-hoc network and cognitive radio.

Each Tx is equipped with $N_t$ antennas and the Rx with $N_r$ antennas. Each Tx communicates with a unique Rx at a time.  Txs are not allowed or able to exchange users' packet (message) information, giving rise to an interference channel over which we seek some form of beamforming-based coordination.  The channel from Tx $i$ to Rx $j$
 $\mathbf{H}_{ji} \in \mathcal{C}^{N_r \times N_t}$ is given by:
 \begin{equation}\label{eqt:channel}
 	\mathbf{H}_{ji}= \sqrt{\alpha_{ji}} \bar{\mathbf{H}}_{ji}, \; \; \; i,j = 1, \ldots, N_c
 \end{equation}
 
Each element in channel matrix $\bar{\mathbf{H}}_{ji}$ is an
independent identically distributed complex Gaussian random variable with zero
mean and unit variance and $\alpha_{ji}$ denotes the slow-varying shadowing and pathloss attenuation. $\bar{\mathbf{H}}_{ji}$ is circularly symmetric complex gaussian and the probability density is
\begin{equation}\label{eqt:complex_gau}
 f_{\bar{\mathbf{H}}_{ji}}(\mathbf{H})= \frac{1}{\pi^{N_tN_r}}exp(-Tr \left(\mathbf{H}\mathbf{H}^H \right)).
\end{equation}

\subsection{Limited Channel knowledge}	
\begin{figure}
\begin{center}
 	\includegraphics[width=9cm, height=7cm]{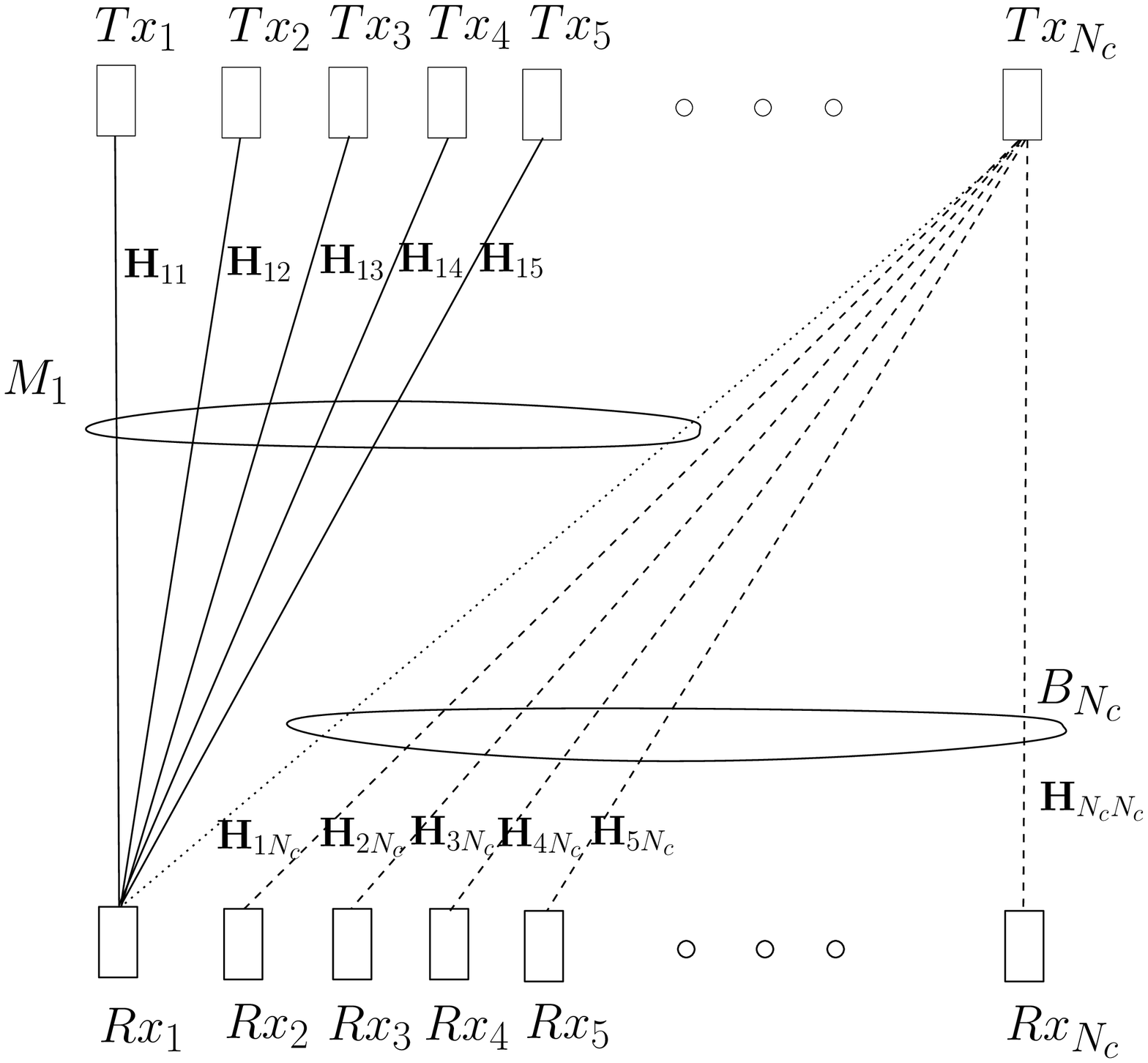}
 	\caption{Limited channel knowledge model: as an illustration, the local CSI available at Tx $N_c$ is shown in dashed lines. The local CSI available at Rx 1 is shown in solid lines.\label{fig:channel_model_bs}}
\end{center}
 \end{figure}

 Although there may exist various ranges and definitions of local CSI, we assume a standard definition of a quasi-distributed CSI scenario where the devices (Tx and Rx alike) are able to  gain knowledge of those local channel coefficients \emph{directly connected} to them, as illustrated in Fig. \ref{fig:channel_model_bs}, possibly complemented with some limited non local information (to be defined later). 

The set of CSI locally available (resp. not available) at Tx $i$ denoted by $\mathbb{B}_i$ (resp. $\mathbb{B}_i^\perp$) is denoted by:
\begin{equation}\label{eqt:ChBS}
	\mathbb{B}_i= \left\{ \mathbf{H}_{ji} \right\}_{j=1, \ldots, N_c} \; ;
\; \mathbb{B}_i^\perp= \left\{ \mathbf{H}_{kl}\right\}_{k,l=1 \ldots N_c}
\setminus \mathbb{B}_i 
\end{equation}
Similarly,  define the set of channels known (resp. unknown) at Rx $i$ denoted by $\mathbb{M}_i$ (resp. $\mathbb{M}_i^\perp$) as:
$
	\mathbb{M}_i= \left\{ \mathbf{H}_{ij} \right\}_{j= 1, \ldots, N_c} \; ;
\; \mathbb{M}_i^\perp = \left\{ \mathbf{H}_{kl}\right\}_{k,l=1 \ldots N_c}
\setminus \mathbb{M}_i .
$ By construction here, locally available channel knowledge, $\mathbb{B}_i$, is only known to Tx $i$ but \emph{not} other Txs. We call this knowledge $\mathbb{B}_i$ the \emph{type} of player (Tx) $i$, in the game theoretic terminology \cite{Harsanyi}.

In the view of Tx $i$, the decision to be made shall be based on its type $\mathbb{B}_i$ and its \emph{beliefs} on other Txs types. Since Tx $i$ does not know other Txs types, we assume that Tx $i$ has a probability density over the possible values of other players channel knowledge $\mathbb{B}_j$. For simplicity, we assume that these \emph{beliefs} are symmetric: the probability density of the \emph{gaussian channels} available at Tx $i$ regarding $\mathbb{B}_j$ is the same as the probability density of Tx $j$ over $\mathbb{B}_i$. The asymmetric path loss antennuations $\alpha_{ji}$ are assumed to be long term satistics and known to the Txs. And we assume that the channel coefficients in the network are statistically independent from each other. We define here the joint beliefs (probability density) at Tx $i$:
\begin{equation}\label{eqt:belief}
 \mu_i= p(\mathbb{B}_i^{\perp})= f_{\bar{\mathbf{H}}_{ji}}(\mathbf{H})^{N_c(N_c-1)}=\mu.
\end{equation} The Tx index $i$ is dropped because the beliefs are symmetric among Txs, given the asymetric path loss coefficients $\alpha_{ji}$. $p(.)$ is a probability measure and $ f_{\bar{\mathbf{H}}_{ji}}(\mathbf{H})$ is the probability  density of a complex gaussian channel defined in \eqref{eqt:complex_gau}.
The second equality relies on the assumptions that the channel coefficients from any Tx to any Rx are independent. 

Based on its \emph{belief}, Tx $i$  designs
the transmit beamforming vector, $\mathbf{w}_i \in \mathcal{C}^{N_t
\times 1}$. As in several important contributions dealing with coordination on the interference channel  
\cite{Randa2008,  Ku2008, Choi2006, Jorswieck2008,  Jafar2008, Shi2008, Farrokhi1998}, we assume linear beamforming. We call the transmit
beamforming vector $\mathbf{w}_i$ an action of Tx $i$ and denote the set of all possible actions by $\mathcal{A}$ at any Tx.
\begin{equation}\label{eqt:userset}
 \mathcal{A}=\left\{ \mathbf{w} \in \mathcal{C}^{N_t \times 1}: \|\mathbf{w}\|^2 \leq 1 \right\}
\end{equation}

The received signal at Rx $i$ is therefore
\begin{equation}
y_i= \mathbf{v}_i^H \mathbf{H}_{ii} \mathbf{w}_i +  \sum_{j \neq i}^{N_c} \mathbf{v}_i^H \mathbf{H}_{ij} \mathbf{w}_j + n_i
\end{equation} where $n_i$ is a gaussian noise with power $\sigma_i^2$. Note that the noise levels $\sigma_i^2$ depend on the link index which was not considered in previous work on transmitter coordination. The Rxs are assumed to employ maximum SINR (Max-SINR) beamforming throughout the paper so as to also maximize the link rates \cite{Paulraj2003}. The receive beamformer $\mathbf{v}_i$ is classically given by:
\begin{equation}\label{eqt:max_sinr}
\mathbf{v}_i= \frac{{\mathbf{C}_{Ri}}^{-1} \mathbf{H}_{ii}
\mathbf{w}_i}{\|{\mathbf{C}_{Ri}}^{-1} \mathbf{H}_{ii} \mathbf{w}_i \|}
\end{equation} where $\mathbf{C}_{Ri}$ is the covariance matrix of received interference and
noise
\begin{equation}\label{eqt:int_noise_pow}
	\mathbf{C}_{Ri}= \sum_{j \neq i} \mathbf{H}_{ij} \mathbf{w}_j \mathbf{w}_j^H
\mathbf{H}_{ij}^H P+ \sigma_i^2 \mathbf{I}.
\end{equation} $P$ is the transmit power. Note that the receive beamformer $\mathbf{v}_i$ is a function of all transmit beamforming vectors $\mathbf{w}_i$. When the transmit beamforming vector $\mathbf{w}_i$ is optimized, the received beamforming vector is modified accordingly.

\begin{figure}
\begin{center}
 \includegraphics[width=9cm, height=7cm]{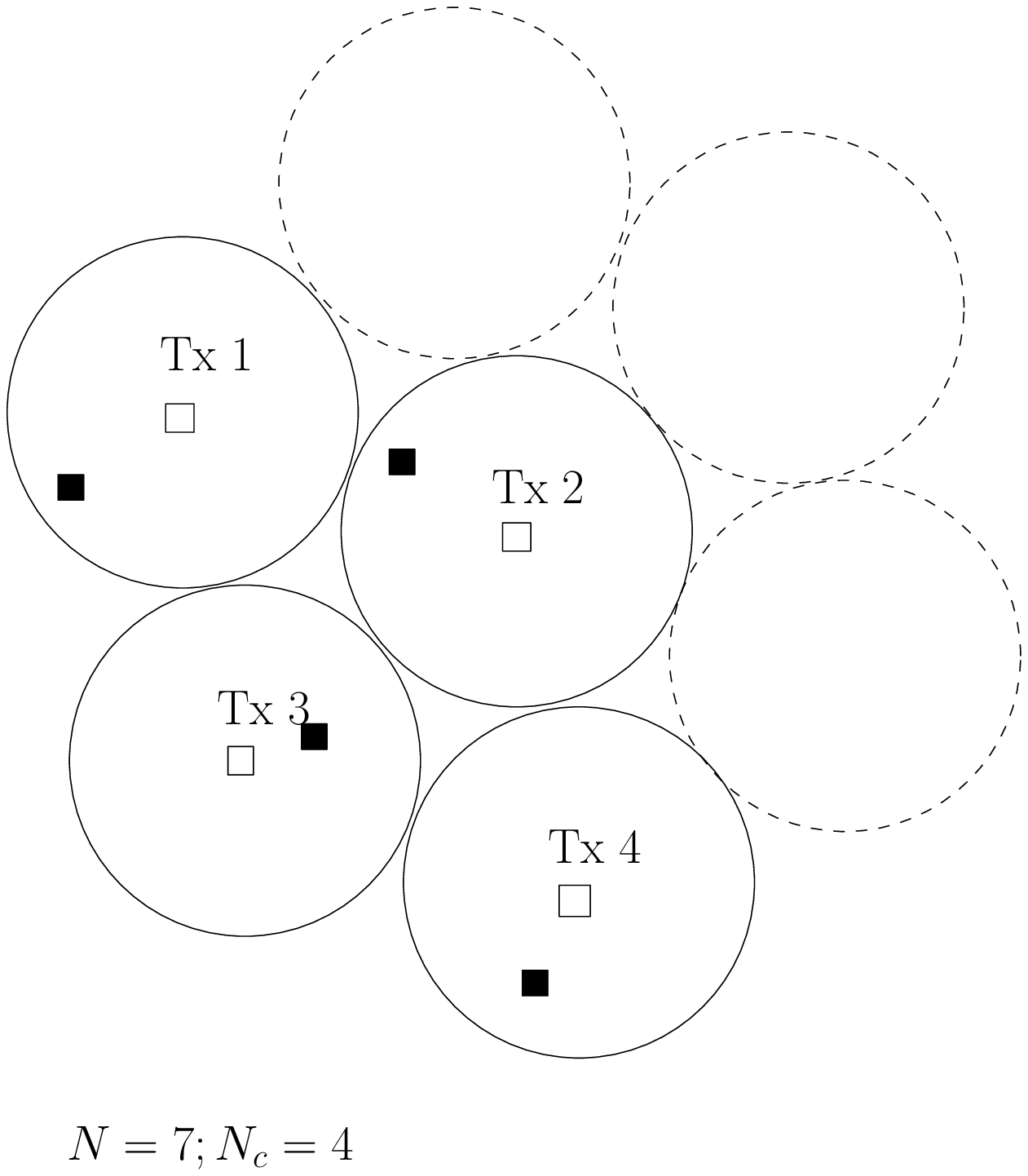}
\caption{This figure illustrates a system of $N=7$ cells where $N_c=4$  form a coordination cluster. Empty squares represent transmitters whereas filled squares represent receivers. The noise power (which includes out of cluster interference) undergone in each cell varies from link to link.} \label{fig:asym_ch_model}
\end{center}
\end{figure}

Importantly, the noise will in practice capture thermal noise effects but also any interference originating from the rest of the network, i.e. coming from transmitters located beyond the coordination cluster. Thus, depending on path loss and shadowing effects, the $ \{\sigma_{i}^2 \}$  may be quite different from each other \cite{molisch}. Fig. \ref{fig:asym_ch_model} illustrates a system of $N=7$ cells where $N_c=4$ form a coordination cluster. Note that we consider the sum of uncoordinated source of interference and thermal noise to be spatially white. The non-colored interference assumption is justified in the scenario where receivers cannot obtain specific knowledge of the interference covariance and can be interpreted as a worst case scenario, since the receivers cannot use their spatial degrees of freedom to further cancel uncontrolled interference.

\emph{Receiver feedback v.s. Reciprocal Channel}:
In the case of reciprocal channels , e.g. time-devision-duplex systems (TDD), the feedback requirement to obtain $\mathbb{B}_i$ can be replaced by a channel estimation step based on uplink pilot sequences. Additionally, it will be classically assumed that the receivers are able to estimate the covariance matrix of their interference signal, based on, say, transmit pilot sequences.

%% file: bayesian_games.tex
We can now define the Bayesian game on interference channel.  
 \begin{Def} The Bayesian game on interference channel can be described by a 5-tuple:
\begin{equation}\label{eqt:game}
 	G= < \mathcal{N}_c, \mathcal{A},\left\{ \mathbb{B}_i\right\},\mu, \left\{ u_i\right\}>
\end{equation}
where $ \mu$ denotes the \emph{beliefs} of the players and $\left\{ u_i \right\}$ denotes
the utility functions of the players, which can be either egoistic or altruistic.
\end{Def} Specific definitions of $u_i$ will be given in the following sections. The players are assumed to be \emph{rational}
as they maximize their own utility based on their \emph{types} and \emph{beliefs}.

\begin{Def}
	A pure-strategy of player $i$, $s_i: \mathbb{B}_i \rightarrow \mathcal{A}_i$ is a
deterministic choice of action given information $\mathbb{B}_i$ of player $i$. 
\end{Def}
\begin{Def}
	A strategy profile $\mathbf{s}^*=(s_i^*, s_{-i}^*)$
achieves the Bayesian Equilibrium if $s_i^*$ is the best response of player $i$ given
strategy tuple $s_{-i}^*$ for all other players and is characterized by
\begin{equation}
	\forall i \; \;	s_i^*=\arg \max
\mathcal{E}_{\mathbb{B}_i^\perp} \left\{ u_i(s_i, s_{-i}^*) \right\} .
\end{equation}
\end{Def}
Note that, intuitively, the player's strategy is optimized by averaging over the \emph{beliefs} (the distribution of all missing state information) while in a standard game, such expectation is not required.

In the following sections, we derive the equilibria for egoistic and altruistic bayesian games respectively. These equilibria constitute extreme strategies which do not perform optimally in terms of the overall network performance, yet can be exploited as components of a more general beamforming-based coordination technique which is then proposed in section \ref{section:dba}.

%% file: egosol_fb.tex
\section{Bayesian Games with Receiver Beamformer Feedback}
We assume that Tx has the local channel state information $\mathbb{B}_i$ and the added
knowledge of receive beamformers through a feedback channel. Note that in the case of reciprocal channels, the receive beamformer feedback is \emph{not} required.

\subsection{Egoistic Bayesian Game}

\begin{Def}\label{def:ego_fb}
	Denote the set of transmit beamforming vectors of players $j, j\neq i$, by $\mathbf{w}_{-i}$. The egoistic utility function for Tx $i$ is defined as its received SINR
\begin{equation}\label{eqt:ego_fb}
 u_i(\mathbf{w}_i,
\mathbf{w}_{-i})= \frac{ |\mathbf{v}_i^H \mathbf{H}_{ii}
\mathbf{w}_i|^2 P}{\sum_{j\neq i}^{N_c} |\mathbf{v}_i^H \mathbf{H}_{ij} \mathbf{w}_j|^2 P
+\sigma_i^2}.
\end{equation} Based on Tx $i$'s belief, Tx $i$ maximizes the utility function in \eqref{eqt:ego_fb} where $\mathbf{v}_i$ is a known quantity. 
\end{Def}

\begin{Lem}
 There exist at least one Bayesian Equilibrium in the egoistic Bayesian Game $G$ \eqref{eqt:game} with utility function defined in \eqref{eqt:ego_fb}. 
\end{Lem}
\begin{proof}
 $\mathcal{A}_i$ is convex, closed and bounded for all players $i$ and the egoistic utility function $u_i(\mathbf{w}_i,\mathbf{w}_{-i})$ is continuous in both $\mathbf{w}_i$ and $\mathbf{w}_{-i}$. The utility function is convex in $\mathbf{w}_i$ for any set $\mathbf{w}_{-i}$. Thus, at least one Bayesian Equilibrium exists \cite{Osborne1994,He2009}. 
\end{proof}

\begin{Thm}
The best-response strategy of player $i$ in the egoistic Bayesian Game $G$ \eqref{eqt:game} with utility function \eqref{eqt:ego_fb} is to maximize the utility function based on 
its belief:
\begin{equation}
 \mathbf{w}_i^{Ego} = \arg \max \mathcal{E}_{\mathbb{B}_i^\perp} \left\{ u_i(\mathbf{w}_i, \mathbf{w}_{-i}) \right\}.
\end{equation}
The best-response strategy of player $i$ is
\begin{equation}
	\mathbf{w}_i^{Ego}= V^{(max)}(\mathbf{E}_i)
\end{equation}
where $\mathbf{E}_i$  denotes the {\em egoistic equilibrium matrix} for Tx $i$, given by 
$$
\mathbf{E}_i=\mathbf{H}_{ii}^H \mathbf{v}_i \mathbf{v}_i^H \mathbf{H}_{ii}.
$$
\end{Thm}
\begin{proof}
The knowledge of receive beamformers decorrelates the maximization problem which can be written as
\begin{eqnarray}
\nonumber\mathbf{w}_i^{Ego} &=&\arg \max_{\|\mathbf{w}_i\|
\leq 1} \mathcal{E}_{\mathbb{B}_i^\perp} \left\{\frac{1}{\sum_{j\neq i}^{N_c}
 |\mathbf{v}_i^H \mathbf{H}_{ij} \mathbf{w}_j|^2 P +\sigma_i^2} \right\} \\
&&  \mathbf{w}_i^H \mathbf{H}_{ii}^H \mathbf{v}_i \mathbf{v}_i^H \mathbf{H}_{ii} \mathbf{w}_i
\end{eqnarray}
The egoistic-optimal transmit beamformer is therefore the dominant eigenvector of $\mathbf{H}_{ii}^H \mathbf{v}_i \mathbf{v}_i^H \mathbf{H}_{ii}$.
\end{proof}

%% file: altsol_fb.tex
\subsection{Altruistic Bayesian Game}
\begin{Def}
	The utility of the altruistic game is defined here so as to minimize the  sum of interference powers caused to other receivers. 
\begin{equation}\label{def:alt}
	u_i(\mathbf{w}_i, \mathbf{w}_{-i})= -\sum_{j \neq
i}|\mathbf{v}_j^H \mathbf{H}_{ji} \mathbf{w}_i|^2	
\end{equation}
\end{Def}
Note that the receive beamforming vectors $\mathbf{v}_j$ is a Max-SINR receiver which depends on the transmit beamforming vectors $\mathbf{w}_j$ and cause conflicts between Txs.
\begin{Lem}
 There exist at least one Bayesian Equilibrium in the altruistic Bayesian Game $G$ \eqref{eqt:game} with utility function defined in \eqref{def:alt}. 
\end{Lem}
\begin{proof}
 $\mathcal{A}_i$ is convex, closed and bounded for all players $i$ and the altruistic utility function $u_i(\mathbf{w}_i,\mathbf{w}_{-i})$ is continuous in both $\mathbf{w}_i$ and $\mathbf{w}_{-i}$. The utility function is concave in $\mathbf{w}_i$ for any set $\mathbf{w}_{-i}$. Thus, at least one Bayesian Equilibrium exists \cite{Osborne1994,He2009} .
\end{proof}

\begin{Thm}
Based on belief $\mu$, Tx $i$ seeks to maximize the utility function defined in \eqref{def:alt}. The
best-response strategy is
\begin{equation}
	\mathbf{w}_i^{Alt}= V^{(min)}(\sum_{j \neq i} \mathbf{A}_{ji})
\label{eqt:mmse}	
\end{equation}
where $\mathbf{A}_{ji}$ denotes the {\em altruistic equilibrium matrix for Tx $i$ towards Rx $j$}, defined by $\mathbf{A}_{ji}=\mathbf{H}_{ji}^H \mathbf{v}_j \mathbf{v}_j^{H} \mathbf{H}_{ji}$. 
\end{Thm}

\begin{proof}
 Recall the utility function to be $-\sum_{j \neq i} |\mathbf{v}_j^H \mathbf{H}_{ji} \mathbf{w}_i|^2= -\sum_{j \neq i} \mathbf{w}_i^H \mathbf{A}_{ji} \mathbf{w}_i$. Since $\mathbf{v}_j$ are known from feedback or estimation in reciprocal channels, the optimal $\mathbf{w}_i$ is the least dominant eigenvector of the matrix $\sum_{j \neq i}\mathbf{A}_{ji}$.
\end{proof}

%% file: dbs_fb.tex
\section{Sumrate Maximization with Receive Beamformer Feedback}
From the results above, it can be seen that balancing altruism and egoism for player $i$ can be done by trading-off between setting the beamformer close to the dominant eigenvectors of the egoistic equilibrium $\mathbf{E}_i$ or that of the negative altruistic equilibrium  $\left\{ -\mathbf{A}_{ji} \right\}$ ($j \neq i$) matrices in \eqref{eqt:mmse}.
Interestingly, it can be shown that sum rate maximizing precoding for the MIMO-IC does exactly that. Thus we hereby briefly re-visit rate-maximization approaches such as \cite{Wolfgang2009} with this perspective. 

Denote the sum rate by $\bar{R}=\sum_{ i=1}^{N_c} R_i$ where $	R_i = \log_2 \left(1+ \frac{|\mathbf{v}_i^H \mathbf{H}_{ii}
\mathbf{w}_i|^2 P }{\sum_{j \neq i}^{N_c} |\mathbf{v}_i^H
\mathbf{H}_{ij} \mathbf{w}_j|^2 P + \sigma_i^2} \right)$.
\vspace{1cm}

\begin{Lem}\label{Lem:opt_lambda}
	The transmit beamforming vector which maximizes the sum rate $\bar{R}$ is the dominant eigenvector of a matrix, which is a
linear combination of $\mathbf{E}_i$ and $\mathbf{A}_{ji}$:
\begin{equation}\label{eqt:ego_alt_linear}
	\left(\mathbf{E}_i + \sum_{j \neq i}^{N_c} \lambda_{ji}^{opt} \mathbf{A}_{ji}
\right) \mathbf{w}_i= \mu_{max} \mathbf{w}_i
\end{equation}  where 
\begin{equation}\label{eqt:opt_lambda}
	\lambda_{ji}^{opt}=-\frac{S_{jj}}{\sum_{k=1}^{N_c} S_{jk}  + \sigma_j^2} \frac{\sum_{k=1}^{N_c} S_{ik}  + \sigma_i^2}{\sum_{k \neq j}^{N_c} S_{jk} + \sigma_j^2}
\end{equation} where $S_{jk}= |\mathbf{v}_j^H \mathbf{H}_{jk}
\mathbf{w}_k|^2 P$ and $\mu_{max}$ is defined in the proof.
\end{Lem}

\begin{proof}
see appendix \ref{app:opt_lam}. 
\end{proof}
Note that the balancing between altruism and egoism in sum rate maximization is done using the dominant eigenvector of a simple {\em linear combination} of the altruistic and egoistic equilibrium matrices.
The balancing parameters, $ \{\lambda_{ji}^{opt} \}$, can be shown simply to coincide with the pricing parameters invoked in the iterative algorithm proposed in \cite{Wolfgang2009}.
Clearly, these parameters plays a key role, however their computation  is a function of the {\em global} channel state information and requires additional message (price) exchange.
 Instead, we seek below a suboptimal egoism-altruism balancing technique which only requires statistical channel information, while exhibiting the right performance scaling when SNR grows large.  
 
\section{A practical distributed beamforming algorithm: \emph{DBA}}\label{section:dba}
We are proposing the following distributed beamforming algorithm (\emph{DBA}) where one computes the transmit and receive beamformers iteratively as:
\begin{eqnarray}\label{eqt:w_lb}
\mathbf{w}_i &=& V^{max} \left(\mathbf{E}_i + \sum_{j \neq i}^{N_c}  \lambda_{ji} \mathbf{A}_{ji} \right) \\
\mathbf{v}_i&=&\frac{\mathbf{C}_{Ri}^{-1}\mathbf{H}_{ii}\mathbf{w}_i}{\|\mathbf{C}_{Ri}^{-1}\mathbf{H}_{ii}\mathbf{w}_i \|}\label{eqt:dbs_rx}
\end{eqnarray}
where $\lambda_{ji}$ shall be made to depend on channel statistics only. At this stage, it is interesting to compare with previous schemes based on interference alignment such as the practical algorithms proposed in \cite{Heath2009}. In such schemes, the transmit beamformer $\mathbf{w}_i$ is taken independent of $\mathbf{H}_{ii}$. Note that here however, $\mathbf{w}_i$ is correlated to the direct channel gain $\mathbf{H}_{ii}$ through the Egoistic matrix $\mathbf{E}_i$ in \emph{DBA}. The correlation is useful in terms of sum rate as it allows proper weighting between the contributions of the egoistic and altruistic matrices in a link specific manner. 

\subsection{The egoism-altruism balancing parameters $\lambda_{ji}$}
\label{app:subopt_lam}
The egoism-altruism balancing parameters $\lambda_{ji}$ are now found heuristically based on the statistical channel information. Recall from \eqref{eqt:opt_lambda} that
\begin{equation}
	\lambda_{ji}^{opt}=-\frac{S_{j}}{S_j + I_j  + \sigma_j^2} \frac{S_i +I_i + \sigma_i^2}{I_j +\sigma_j^2}
\end{equation} where $S_j=|\mathbf{v}_j^H \mathbf{H}_{jj}
\mathbf{w}_j|^2 P$ and $I_j=\sum_{k \neq j}^{N_c} |\mathbf{v}_j^H \mathbf{H}_{jk}
\mathbf{w}_k|^2 P$. 

Following the principle behind sum rate maximization, we conjecture that at convergence, residual coordinated interference shall be proportionate to the noise
and out-of-cluster interference, i.e. $I_j = O(\sigma_j^2).$ Note that this should not be interpreted as an assumption in a proof but rather as a proposed design guideline. Based on this, we propose the following characterization:

\begin{equation}
	\lambda_{ji}^{opt}=-\frac{S_j}{S_j + O(\sigma_j^2)} \frac{S_i + O( \sigma_i^2)}{O(\sigma_j^2)}.
\end{equation}
Note that $S_j$ and $S_i$ are independent and we have
\begin{eqnarray}
\lambda_{ji}^{opt}&=& - \frac{S_j}{S_j + O(\sigma_j^2)} \frac{S_i + O(\sigma_i^2)}{O(\sigma_j^2)}\\
\mathcal{E} \lambda_{ji}^{opt} &\overset{(a)}{=}& - \mathcal{E} \left(\frac{S_j}{S_j + O(\sigma_j^2)}  \right)  \frac{\mathcal{E} (S_i) + O(\sigma_i^2)}{O(\sigma_j^2)}\\
 &\overset{(b)}{\geq}& -  \frac{\mathcal{E} S_j}{\mathcal{E} S_j + O(\sigma_j^2)}   +\frac{\mathcal{E} (S_i) + O(\sigma_i^2)}{O(\sigma_j^2)}\\
 &=&  -  \frac{1}{1 + \frac{ O(\sigma_j^2)}{\mathcal{E} S_j}}    \frac{1 + \frac{O(\sigma_i^2)}{\mathcal{E} (S_i)} }{ \frac{O(\sigma_j^2)}{\mathcal{E} (S_i)}}
\end{eqnarray}
where $(a)$ is because $S_i,S_j$ are independent and $(b)$ is because the function $\frac{x}{x+c}$ is concave in $x$ and therefore by Jensen's inequality, we have $\frac{\mathcal{E} x}{\mathcal{E} x+c} \geq \mathcal{E} \frac{x}{x+c}$. 

Although $\mathcal{E} S_i$ is not known explicitly, it is strongly  related to the strength of the direct channel $P \alpha_{ii}$. Let $\gamma_i= \frac{P \alpha_{ii}}{\sigma_i^2}$. In order to obtain an exploitable formulation for $\lambda_{ji}$, we replace $\mathcal{E} S_i$ by $P \alpha_{ii}$ and $O(\sigma_i^2)$ by $\sigma_i^2$, to derive:
\begin{equation}\label{eqt:lambda}
	\lambda_{ji}  = -\frac{1}{1 +\gamma_j^{-1}} \frac{1 + \gamma_i^{-1}}{\frac{\sigma_j^2}{P \alpha_{ii}}}.
\end{equation}
Interestingly, in the special case where direct channels have the same average strength, we obtain a simple expression
\begin{equation}
	\lambda_{ji} = -\frac{1+ \gamma_i^{-1}}{1 +\gamma_j^{-1}} \gamma_j.
\end{equation} 
The above result suggests Tx $i$ to behave more altruistically towards link $j$ when the SNR of link $j$ is high or when the SNR of link $i$ is comparatively lower. This is in accordance with the intuition behind rate maximization over parallel gaussian channels.

\emph{DBA} iterates between optimizing the transmit and receive beamformers, as summarized in Algorithm \ref{fig:dba_algo}. Iterating between transmit and receive beamformers is reminiscent of  recent interference-alignment based methods  \cite{Jafar2008,Heath2009}. However here, interference alignment is {\em not} a design criterion. In  \cite{Jafar2008}, an improved interference alignment technique based on alternately maximizing the SINR at both transmitter and receiver sides is proposed. In contrast, here the Max-SINR criterion is only used at the receiver side. Although the distinction is unimportant in the large SNR case (see below), it dramatically changes performance in certain situations at finite SNR (see Section \ref{section:result}).
\begin{algorithm}
\caption{DBA \label{fig:dba_algo}}
\begin{algorithmic}
 \STATE 1) Initialize beamforming vectors $\mathbf{w}_i, i=1, \ldots, N_c$, to be predefined vectors.
\STATE 2) For each Rx $i$, compute $\mathbf{v}_i = \frac{\mathbf{C}_{Ri}^{-1} \mathbf{H}_{ii} \mathbf{w}_i}{\|\mathbf{C}_{Ri}^{-1} \mathbf{H}_{ii} \mathbf{w}_i\|}$
where $\mathbf{C}_{Ri}$ is computed with $\mathbf{w}_i$ in previous step.
\STATE 3) For each Tx $i$, compute $\mathbf{w}_i = V^{max} \left(\mathbf{E}_i + \sum_{j \neq i}^{N_c}  \lambda_{ji} \mathbf{A}_{ji} \right)$
where $\lambda_{ji}$ are computed from satistical parameters \eqref{eqt:lambda}.
\STATE 4) Repeat step 2 and 3 until convergence.
\end{algorithmic}
\end{algorithm}
\subsection{Asymtotic Interference Alignment}
One important aspect of the algorithm above is whether it achieves the interference alignment in high SNR regime \cite{Jafar2008}. The following theorem answers this question positively.

\begin{Def}
Define the set of beamforming vectors solutions in downlink (respectively uplink) interference alignment to be \cite{Jafar2008}
\begin{eqnarray}
\nonumber \mathcal{IA}^{DL} &=& \label{def:low_rank} \left\{ (\mathbf{w}_1, \ldots, \mathbf{w}_{N_c}): \right.\\
 && \left. \sum_{k \neq i}^{N_c} \mathbf{H}_{ik} \mathbf{w}_k \mathbf{w}_k^{H} \mathbf{H}_{ik}^H \text{ is low rank, } \forall i \right\}\\
\nonumber \mathcal{IA}^{UL} &=& \left\{ (\mathbf{v}_1, \ldots, \mathbf{v}_{N_c}): \right. \\
&& \left.\sum_{k \neq i}^{N_c} \mathbf{H}_{ki}^H \mathbf{v}_k \mathbf{v}_k^{H} \mathbf{H}_{ki} \text{ is low rank, } \forall i \right\}.
\end{eqnarray}
Thus, for all $(\mathbf{w}_i,\ldots, \mathbf{w}_{N_c}) \in \mathcal{IA}^{DL}$, there exist receive beamformers $\mathbf{v}_i, i=1,\ldots,N_c$ such that the following is satisfied:
\begin{equation}\label{def:IA}
\mathbf{v}^H_i \mathbf{H}_{ij} \mathbf{w}_j =0 \; \; \forall i,j \neq i .
\end{equation}
\end{Def} 

Note that the uplink alignment solutions are defined for a virtual uplink having the same frequency and only appear here as a technical concept helping with the proof. 

\begin{Thm}\label{thm:converge}
Assume the downlink interference alignment set is non-empty (interference alignment is feasible). Denote average SNR of link $i$ by $\gamma_i=\frac{P \alpha_{ii}}{\sigma_i^2}$. Let $ \lambda_{ji} = -\frac{1+\gamma_i^{-1}}{1+\gamma_j^{-1}}\gamma_j$, then in the large SNR regime, $P \rightarrow \infty$ , any transmit beamforming vector in $\mathcal{IA}^{DL}$ is a convergence (stable) point of \emph{DBA}.
\end{Thm}
\begin{proof}
  see Appendix \ref{appendix:lambda_IA}.
\end{proof}

Note that this does not prove global convergence, but local convergence, as is the case for other IA or rate maximization techniques \cite{Jafar2008, Heath2009, Wolfgang2009}. Another way to characterize local convergence is as follows: assuming interference alignment is feasible ($\mathcal{IA}^{DL}$ is non-empty), the first algorithm in \cite{Jafar2008} was shown to converge to transmit beamformers belonging to $\mathcal{IA}^{DL}$  and the receivers are based on the minimum eigenvector of the dowlink interference covariance matrix, which tends to be low-rank. However,  \emph{DBA} selects its receive beamformer from the Max-SINR criterion which, in the large SNR situation, is also identical to selecting receive beamformers in the null space of the interference covariance matrix. Therefore when interference alignment is feasible, the algorithm in \cite{Jafar2008} and \emph{DBA} coincide at large SNR. This aspect is confirmed by our simulations (see section \ref{section:result}).

%% file: appendix.tex
\section{Appendix}

\subsection{Proof of Lemma \ref{Lem:opt_lambda}}
\label{app:opt_lam}
 Define the largrangian of the sum rate maximization problem for Tx $i$ to be 
$\mathcal{L}(\mathbf{w}_i ,\mu)= \bar{R} - \mu_{max}( \mathbf{w}_i^H \mathbf{w}_i -1)$. The neccessary condition of largrangian $\frac{\partial}{\partial
\mathbf{w}_i^H}\mathcal{L}(\mathbf{w}_i ,\mu)=0$ gives:
$\frac{\partial}{\partial \mathbf{w}_i^H} R_i + \sum_{j \neq i}^{N_c}
\frac{\partial}{\partial \mathbf{w}_i^H} R_j = \mu_{max} \mathbf{w}_i 
$. With elementary matrix calculus, 

\begin{eqnarray}
&&\frac{\partial}{\partial \mathbf{w}_i^H} R_i  =
\frac{P}{\sum_{k=1}^{N_c} |\mathbf{v}_i^H \mathbf{H}_{ik}
\mathbf{w}_k|^2 P + \sigma_i^2} \mathbf{E}_i \mathbf{w}_i\\
\nonumber &&\frac{\partial}{\partial \mathbf{w}_i^H} R_j = -\frac{|\mathbf{v}_j^H \mathbf{H}_{jj} \mathbf{w}_j|^2 P}{\sum_{k=1}^{N_c} |\mathbf{v}_j^H \mathbf{H}_{jk}
\mathbf{w}_k|^2 P  + \sigma_j^2}\\
&&  \frac{P }{\sum_{k \neq
j}^{N_c} |\mathbf{v}_j^H \mathbf{H}_{jk} \mathbf{w}_k|^2 P +
\sigma_j^2} \mathbf{A}_j \mathbf{w}_i
\end{eqnarray}

 where $\lambda_{ji}^{opt}$ is a function of all channel states information and
beamformer feedback:
\begin{equation}
	\lambda_{ji}^{opt}=-\frac{S_{jj}}{\sum_{k=1}^{N_c} S_{jk}  + \sigma_j^2} \frac{\sum_{k=1}^{N_c} S_{ik}  + \sigma_i^2}{\sum_{k \neq j}^{N_c} S_{jk} +\sigma_j^2}
\end{equation} where $S_{jk}=|\mathbf{v}_j^H \mathbf{H}_{jk}
\mathbf{w}_k |^2 P$.
Thus, the gradient is zero for any $\mathbf{w}_i$ eigenvector of the matrix shown on the L.H.S. of \eqref{eqt:ego_alt_linear}.  Among all stable points, the global maximum of the cost function is reached by selecting the dominant eigenvector of $E_i + \sum_{j \neq i} \lambda_{ji} A_{ji}$ .

\subsection{Proof of Theorem \ref{thm:converge}: convergence points of \emph{DBA}}\label{appendix:lambda_IA}

To prove that interference alignment forms a convergence set of \emph{DBA}, we will prove that if  \emph{DBA} achieves interference alignment, \emph{DBA} will not deviate from the solution (stable point).

Assumed interference alignment is reached and let $(\mathbf{w}_1^{IA}, \ldots, \mathbf{w}_{N_c}^{IA}) \in \mathcal{IA}^{DL}$ and $(\mathbf{v}_1^{IA}, \ldots, \mathbf{v}_{N_c}^{IA}) \in \mathcal{IA}^{UL}$. Let $\mathbf{Q}_i^{DL}=\sum_{k \neq i}^{N_c} \mathbf{H}_{ik} \mathbf{w}_k^{IA} \mathbf{w}_k^{IA,H} \mathbf{H}_{ik}^H$ and $\mathbf{Q}_i^{UL}=\sum_{k \neq i}^{N_c} \mathbf{H}_{ki}^H \mathbf{v}_k^{IA} \mathbf{v}_k^{IA,H} \mathbf{H}_{ki}$.

Given receivers $(\mathbf{v}_1^{IA}, \ldots, \mathbf{v}_{N_c}^{IA})$, we compute new transport beamformers. In high SNR regime, $\lambda_{ji} \rightarrow - \infty$  and \emph{DBA} gives $\mathbf{w}_i = V^{min}( \mathbf{Q}_i^{UL})$ \eqref{eqt:w_lb}. By  \eqref{def:low_rank}, $\mathbf{Q}_i^{UL}$ is low rank and thus $\mathbf{w}_i$ is in the null space of $\mathbf{Q}_i^{UL}$. In direct consequence, the conditions of interference alignment \eqref{def:IA} are satisfied. Thus, $(\mathbf{w}_1, \ldots, \mathbf{w}_{N_c}) \in \mathcal{IA}^{DL}$.

Given transmitters $(\mathbf{w}_1^{IA}, \ldots, \mathbf{w}_{N_c}^{IA})$, we compute new receive beamformers.  The receive beamformer is defined as $\mathbf{v}_i=\arg \max \frac{\mathbf{v}_i^H \mathbf{H}_{ii} \mathbf{w}_i^{IA} \mathbf{w}_i^{IA,H} \mathbf{H}_{ii}^H \mathbf{v}_i}{\mathbf{v}_i^H \mathbf{Q}_i^{DL} \mathbf{v}_i}$. Since $\mathbf{Q}_i^{DL}$ is low rank, the optimal $\mathbf{v}_i$ is in the null space of $\mathbf{Q}_i^{DL}$. Hence, $\mathbf{v}_i \in \mathcal{IA}^{UL}$.

Since both $\mathbf{w}_i$ and $\mathbf{v}_i$ stays within $\mathcal{IA}^{DL}$ and $\mathcal{IA}^{UL}$, interference alignment is a convergence point of \emph{DBA} in high SNR.